\documentclass
[preprint,prd,secnumarabic,showkeys,showpacs,nofootinbib,hyperref]{revtex4}%
\usepackage{amsfonts}
\usepackage{amsmath}
\usepackage{amssymb}
\usepackage{graphicx}%
\providecommand{\U}[1]{\protect\rule{.1in}{.1in}}
\begin{document}
\title{Noncommutative Complex Scalar Field and Casimir Effect}
\author{Farid Khelili}
\affiliation{20 Aout 55 Skikda University, Skikda, Algeria}

\begin{abstract}
A noncommutative complex scalar field, satisfying the deformed canonical
commutation relations proposed by Carmona et al. \cite{OSC13}-\cite{OSC17}, is
constructed. Using these noncommutative deformed canonical commutation
relations, a model describing the dynamics of the noncommutative complex
scalar field is proposed. The noncommutative field equations are solved, and
the vacuum energy is calculated to the second order in the parameter of
noncommutativity. As an application to this model, the Casimir effect, due to
the zero point fluctuations of the noncommutative complex scalar field, is
considered. It turns out that in spite of its smallness, the noncommutativity
gives rise to a repulsive force at the microscopic level, leading to a
modified Casimr potential with a minimum at the point $a_{\min}=\sqrt{\frac
{5}{84}}\pi\theta$.

\end{abstract}
\maketitle

\section{Introduction}

Since the birth of quantum mechanics and general relativity, many efforts have
been made to understand the nature of spacetime at very short distances of the
order of the Planck length, or at very high energies, there is now a common
belief that the usual picture of spacetime as a smooth pseudo-Riemannian
manifold should breakdown at very short distances of the order of the Planck
length, due to the quantum gravity effects. Several physical arguments are
used to motivate a deviation from the flat-space concept at very short
distances of the order of the Planck length. Among the new concepts are
quantum groups, quantum loop gravity, deformation theories, noncommutative
geometry, noncommutative spacetime etc... The concept of noncommutative
spacetime was suggested very early on by the founding fathers of quantum
mechanics and quantum field theory. This was motivated by the need to remove
the divergences that arise in quantum electrodynamics. However, this
suggestion was ignored \cite{NCS1a}. The concept of noncommutative spacetime
was discovered in string theory and in the matrix model of M theory, where
noncommutative gauge theory appears as a certain limit in the presence of a
background field B \cite{string1}-\cite{string3}.

In recent years, the idea of noncommutative spacetime has attracted
considerable interest, and has penetrated into various fields in physics and
mathematical physics, starting from the Standard Model of particle physics,
strings, renormalization, to the quantum Hall effect, two-dimensional
noncommutative harmonic oscillators, and noncommutative field theory
\cite{NCS1a}-\cite{Saha3}. One of the new features of noncommutative field
theories is the UV/IR mixing phenomenon, in which the physics at high energies
affects the physics at low energies, which does not occur in quantum field
theories in which the coordinates commute \cite{NCS1a} \cite{NCS12}
\cite{NCS13}. The study of noncommutative spacetime and its implications to
gauge and gravity theories, quantum field theories and other area of
theoretical physics, is motivated by the fact that the effects of
noncommutativity of space may appear at very short distances of the order of
the Planck length, or at very high energies, this may shed a light on the real
microscopic geometry and structure of our universe.

In the last few years, there has been a great interest in the Casimir effect,
for both the theoretical and the experimental sides, it finds applications in
various physical phenomena, such as quantum field theory, condensed matter
physics, elementary particle physics, quantum reflection of atoms on different
surfaces and Bose-Einstein condensation \cite{KAMilton} \cite{Bordag}.

In gravitation and cosmology the Casimir effect can drive the inflation
process and leads to interesting effects in brane models of the universe
\cite{Bordag} \cite{inf1}-\cite{inf4}. More practical reasons for the recent
interest in Casimir effects are their implications for nanotechnology
\cite{KAMilton} \cite{Bordag} \cite{inf4}-\cite{inf6}, where the attractive
forces could lead to restrictions in the construction of nanodevices
\cite{Bordag} \cite{braz1}. Casimir forces are usually attractive, but
repulsive Casimir forces can be achieved in special circumstances, repulsive
Casimir forces might prove useful in the construction of nanodevices, and
other systems in which material components are in close proximity, including
quantum levitation, quantum friction etc... \cite{Bordag} \cite{braz1}.

The Casimir effect provides an effective mechanism for spontaneous
compactification of extra spatial dimensions in multi-dimensional physics,
indeed the vacuum fluctuations of higher-dimensional gravitational field may
contribute an attractive Casimir force to push the size of the extra spaces in
Kaluza-Klein unified theory and string theories to the Planck scale. Near the
Planck scale, it is generally believed that the non-perturbative quantum
gravity can stabilize the size of the extra spaces \cite{Bordag} \cite{inf2}
\cite{extra1}-\cite{extra3}. It has been shown that the Casimir energy could
give repulsive force if some of the extra dimensions are noncommutative, this
suggests that the noncommutativity of spatial dimensions provides a possible
mechanism to stabilize the extra radius in high temperature \cite{extra2}%
-\cite{extra6}.

In this paper, we present a model describing a noncommutative complex scalar
field theory with commutative base space \ and noncommutative target space,
and we explore possible implications that the noncommutativity in the target
space might have on the Casimir force. It turns out that in spite of its
smallness, the noncommutativity gives rise to a repulsive force at the
microscopic level, leading to a modified Casimr potential with a minimum, this
result is important, as mentioned above, in nanotechnology and in the
stabilization of the size of the extra spatial dimensions.

By generalizing the noncommutative harmonic oscillator construction\ an
extension of quantum field theory based on the concept of noncommutative
target space has been proposed in \cite{OSC13}-\cite{OSC15}, where the
properties and phenomenological implications of the noncommutative field have
been studied and applied to different problems including scalar, gauge and
fermionic fields \cite{OSC13}-\cite{OSC17}. The idea of noncommutative target
space has also been developed in the work of Balachandran et al. \cite{bala 1}.

Our paper is organized as follows: In Section 2, we consider a noncommutative
action for a complex scalar field with self interaction, in section 3, we
derive and solve the free noncommutative field equations, in section 4, we
\ consider the noncommutative Casimir effect. Finally, in section 5, we draw
our conclusions.

\section{Noncommutative Action}

Consider a complex scalar field $\Phi\left(  x\right)  $ with Lagrangian
density given by \cite{WEINB}-\cite{BRN}%

\begin{equation}
\mathfrak{L}=-\left(  \partial_{\mu}\Phi\right)  ^{\ast}\left(  \partial^{\mu
}\Phi\right)  -m^{2}\Phi^{\ast}\Phi-g\left(  \Phi^{\ast}\Phi\right)  ^{2}
\label{eqn1}%
\end{equation}

where $m$ is the mass of the charged particles, and $g$ is a positive
parameter. The metric signature will be assumed to be $-++$..., in what
follows, we take $\hbar=c=1$.

The complex scalar field can be quantized using the canonical quantization
rules, for this we express it in terms of its real and imaginary parts as
$\Phi=\frac{1}{\sqrt{2}}\left(  \varphi_{1}+i\varphi_{2}\right)  $, where
$\varphi_{1},\varphi_{2}$ are real scalar fields; in terms of these real
scalar fields the Lagrangian density reads%

\begin{equation}
\mathfrak{L}=-\frac{1}{2}\left(  \partial_{\mu}\varphi_{a}\right)  ^{2}%
-\frac{1}{2}\mu^{2}\left[  \varphi\right]  \left(  \varphi_{a}\right)  ^{2}
\label{eqn2}%
\end{equation}

where $\mu^{2}\left[  \varphi\right]  =m^{2}+\frac{1}{2}g\left(  \varphi
_{a}\right)  ^{2}.$

Let $\pi_{a}$ be the canonical conjugate to $\varphi_{a}$%
\begin{equation}
\pi_{a}=\frac{\partial\mathfrak{L}}{\partial\overset{.}{\varphi}_{a}}%
=\overset{\cdot}{\varphi}_{a} \label{eqn3}%
\end{equation}

The Hamiltonian density reads then
\begin{equation}
\mathcal{H}=\pi_{a}\overset{\cdot}{\varphi}_{a}-\mathfrak{L}\mathcal{=}%
\frac{1}{2}\left(  \pi_{a}\right)  ^{2}+\frac{1}{2}\left(  \overrightarrow
{\nabla}\varphi_{a}\right)  ^{2}+\frac{1}{2}\mu^{2}\left[  \varphi\right]
\left(  \varphi_{a}\right)  ^{2} \label{eqn4}%
\end{equation}

where the summation convention over repeated indices is assumed throughout
this paper.

To quantize the system, we split the Hamiltonian density $\mathcal{H=H}%
_{0}\mathcal{+H}_{int}$ into a free and interaction terms \cite{WEINB} \ \
\begin{align}
\mathcal{H}_{0}  &  =\frac{1}{2}\left(  \pi_{a}\right)  ^{2}+\frac{1}%
{2}\left(  \overrightarrow{\nabla}\varphi_{a}\right)  ^{2}+\frac{1}{2}%
m^{2}\left(  \varphi_{a}\right)  ^{2}\\
\mathcal{H}_{int}  &  =\frac{1}{4}g\left(  \varphi_{a}\varphi_{a}\right)  ^{2}%
\end{align}

then we pass to the interaction picture. In the interaction picture the
equation of motion are given by%

\begin{equation}
\overset{\cdot}{\varphi}_{a}\left(  x\right)  =\frac{\delta H_{0}}{\delta
\pi_{a}\left(  x\right)  }\text{ \ , \ }\overset{\cdot}{\pi}_{a}\left(
x\right)  =-\frac{\delta H_{0}}{\delta\varphi_{a}\left(  x\right)  }%
\end{equation}

where $H_{0}=\int d^{3}\overrightarrow{x}\mathcal{H}_{0}$ is the free Hamiltonian.

In the canonical quantization the canonical variables $\varphi_{a}$ and the
canonical conjugates $\pi_{a}$ are assumed to be operators satisfying the
canonical commutation relations%

\begin{align}
\left[  \varphi_{a}\left(  t,\overrightarrow{x}\right)  ,\pi_{b}\left(
t,\overrightarrow{y}\right)  \right]   &  =i\delta_{ab}\delta^{3}\left(
\overrightarrow{x}-\overrightarrow{y}\right) \label{eqn5}\\
\left[  \varphi_{a}\left(  t,\overrightarrow{x}\right)  ,\varphi_{b}\left(
t,\overrightarrow{y}\right)  \right]   &  =0\nonumber\\
\left[  \pi_{a}\left(  t,\overrightarrow{x}\right)  ,\pi_{b}\left(
t,\overrightarrow{y}\right)  \right]   &  =0\nonumber
\end{align}

It is well known, since the birth of quantum field theory in the papers of
Born, Dirac, Fermi, Heisenberg, Jordan, and Pauli, that the free field behaves
like an infinite number of coupled harmonic oscillators \cite{WEINB}, using
this analogy between free fields and an infinite number of coupled harmonic
oscillators, one can impose non commutativity on the configuration space of
dynamical fields $\varphi_{a}$, to do this we recall that the two-dimensional
harmonic oscillator noncommutative configuration space can be realized as a
space where the coordinates $\widehat{x}_{a}$, and the corresponding
noncommutative momentum $\widehat{p}_{a},$ are operators satisfying the
commutation relations%

\begin{equation}
\left[  \widehat{x}_{a},\widehat{x}_{b}\right]  =i\theta^{2}\varepsilon
_{ab}\ \ \ \ \ \left[  \widehat{p}_{a},\widehat{p}_{b}\right]
=0\ \ \ \ \ \left[  \widehat{x}_{a},\widehat{p}_{b}\right]  =i\delta_{ab}
\label{eqn5a}%
\end{equation}

where $\theta$ is a parameter with dimension of length, and $\varepsilon_{ab}$
is an antisymmetric constant matrix.\bigskip

It is well known that this noncommutative algebra can be mapped to the
commutative Heisenberg-Weyl algebra \cite{Saha1}-\cite{Saha3}
\begin{equation}
\left[  x_{a},x_{b}\right]  =0\ \ \ \ \ \left[  p_{a},p_{b}\right]
=0\ \ \ \ \ \left[  x_{a},p_{b}\right]  =i\delta_{ab} \label{eqn5b}%
\end{equation}

through the relations%

\begin{equation}
\widehat{x}_{a}=x_{a}-\frac{1}{2}\theta^{2}\varepsilon_{ab}p_{b}%
\ \ \ \ \ \widehat{p}_{a}=p_{a} \label{eqn5c}%
\end{equation}

To impose non commutativity on the configuration space of dynamical fields
$\varphi_{a},$ let us first recall that, in the language of fiber bundles,
classical matter fields may be regarded as local sections of some associated
vector bundle $\left(  E,M,\pi,F\right)  $ to a $G$-principal fiber bundle
with structure group $G$, whose fibers $E_{x}$ are copies of some finite
dimensional vector space $F$ carrying a representation of the structure group
$G$ \cite{fib1}-\cite{fib3}. In particular the complex scalar field
$\Phi=\left(
\begin{array}
[c]{c}%
\varphi_{1}\\
\varphi_{2}%
\end{array}
\right)  $ may be regarded as a section of a vector bundle $\left(
E,M,\pi,\mathbb{R}^{2}\right)  $ whose fibers $E_{x}$ are copies of the real
plane $\mathbb{R}^{2}$, the base space $M$\ will be assumed to have the
topological structure $M=\mathbb{R}\times\Sigma$, where $\mathbb{R}$ is the
real line and $\Sigma$ is some region (compact or noncompact) of
$\mathbb{R}^{3}$%
\begin{equation}
\Phi:M\rightarrow E\text{ \ ,\ }x\rightarrow\Phi\left(  x\right)  =\left(
\begin{array}
[c]{c}%
\varphi_{1}\left(  x\right) \\
\varphi_{2}\left(  x\right)
\end{array}
\right)  \in E_{x}\equiv\mathbb{R}^{2}%
\end{equation}

To define noncommutative dynamical fields $\widehat{\varphi}_{a}$, we replace
the target space $E,$ whose fibers $E_{x}$ are copies of the real plane
$\mathbb{R}^{2},$ by the target space $\widehat{E},$ whose fibers $\widehat
{E}_{x}$ are copies of the noncommutative plane $\widehat{\mathbb{R}}^{2}$.
The noncommutative plane $\widehat{\mathbb{R}}^{2}\ $is generated by two
elements $\left(  \widehat{q}_{a}\right)  ,$ such that the noncommutative
canonical variables $\widehat{q}_{a}$ and the noncommutative canonical
conjugates $\widehat{p}_{a}$ satisfy the noncommutative algebra
\begin{align}
\left[  \widehat{q}_{a},\widehat{q}_{b}\right]   &  =i\theta\epsilon
_{ab}\text{ \ \ \ }a,b=1,2\label{eqn1ab}\\
\left[  \widehat{q}_{a},\widehat{p}_{b}\right]   &  =i\delta_{ab}\text{
\ ,\ \ \ }\left[  \widehat{p}_{a},\widehat{p}_{b}\right]  =0\nonumber
\end{align}

where $\theta$ is the parameter of noncommutativity, with dimension of length,
which is assumed to be a constant, and $\varepsilon_{ab}$ is a $2\times2$ real
antisymmetric matrix%

\begin{equation}
\varepsilon_{12}=-\varepsilon_{21}=1
\end{equation}

Hence each fiber $\widehat{E}_{x}$ is generated by two elements $\left(
\widehat{q}_{a,x}\right)  $ satisfying the noncommutative algebra $\left(
\ref{eqn1ab}\right)  .$ Now we extend the noncommutative algebra $\left(
\ref{eqn1ab}\right)  $ to the total space%

\begin{align}
\left[  \widehat{q}_{a,x},\widehat{q}_{b,x}\right]   &  =i\theta\epsilon
_{ab}\text{ \ \ \ }a,b=1,2\label{eqn1ac}\\
\left[  \widehat{q}_{a,x},\widehat{p}_{b,x}\right]   &  =i\delta_{ab}\text{
\ , \ \ }\left[  \widehat{p}_{a,x},\widehat{p}_{b,y}\right]  =0\text{
}\nonumber\\
\left[  \widehat{q}_{a,x},\widehat{q}_{b,y}\right]   &  =\text{\ }\left[
\widehat{q}_{a,x},\widehat{p}_{b,y}\right]  =0\text{ \ \ if }x\neq y\nonumber
\end{align}

Here, and in the following, all the commutation relations are at equal time
$x^{0}=y^{0}.$

The noncommutative complex scalar field $\widehat{\Phi}=\left(
\begin{array}
[c]{c}%
\widehat{\varphi}_{1}\\
\widehat{\varphi}_{2}%
\end{array}
\right)  $ is the section of the noncommutative bundle $\left(  \widehat
{E},M,\pi,\widehat{\mathbb{R}}^{2}\right)  $%
\begin{equation}
\widehat{\Phi}:M\rightarrow\widehat{E}\text{ \ ,\ }x\rightarrow\widehat{\Phi
}\left(  x\right)  =\left(
\begin{array}
[c]{c}%
\widehat{\varphi}_{1}\left(  x\right) \\
\widehat{\varphi}_{2}\left(  x\right)
\end{array}
\right)  \in\widehat{E}_{x}\equiv\widehat{\mathbb{R}}^{2}%
\end{equation}

defined by%
\begin{equation}
\widehat{\varphi}_{a}\left(  x\right)  =\frac{\sqrt{\Omega}}{\left(
2\pi\right)  ^{\frac{3}{2}}}\frac{\omega_{ab}\left(  x\right)  }{\sqrt
{\det\omega\left(  x\right)  }}\widehat{q}_{b,x}%
\end{equation}

where $\omega_{ab}\left(  x\right)  $ is a spacetime $2\times2$ real symmetric
matrix such that $\det$ $\left[  \omega_{ab}\left(  x\right)  \right]  >0$ for
all $x\in M$, and $\Omega$ is a constant with dimension of $\left(
\text{length}\right)  ^{-3},$ defined by%
\begin{equation}
\Omega=\int^{\left(  \Lambda\right)  }d^{3}\overrightarrow{k}=4\pi%
%TCIMACRO{\dint \nolimits_{0}^{\Lambda}}%
%BeginExpansion
{\displaystyle\int\nolimits_{0}^{\Lambda}}
%EndExpansion
k^{2}dk
\end{equation}

where the presence of $\left(  \Lambda\right)  $ denotes that the integration
only extends over values of $k$ with $\left\vert k\right\vert $ $\leq\Lambda,$
the introduction of a cut-off $\Lambda$ in the momentum space integration is
motivated by the fact that quantum field theory is just a low energy
approximation of a more fundamental theory. So in the limit $\Lambda
\rightarrow\infty,$ $\Omega$ can be replaced by $\left(  2\pi\right)
^{3}\delta^{3}\left(  0\right)  $%
\begin{equation}
\delta^{3}\left(  0\right)  \equiv\delta^{3}\left(  \overrightarrow
{x}-\overrightarrow{x}\right)  =\int\frac{d^{3}\overrightarrow{k}}{\left(
2\pi\right)  ^{3}}=\frac{1}{2\pi^{2}}%
%TCIMACRO{\dint \nolimits_{0}^{\infty}}%
%BeginExpansion
{\displaystyle\int\nolimits_{0}^{\infty}}
%EndExpansion
k^{2}dk
\end{equation}

Using the noncommutative relations $\left(  \ref{eqn1ac}\right)  $ one can see
that, the noncommutative canonical variables $\widehat{\varphi}_{a}$ satisfy
the noncommutative commutation relations%

\begin{align}
\left[  \widehat{\varphi}_{a}\left(  t,\overrightarrow{x}\right)
,\widehat{\varphi}_{b}\left(  t,\overrightarrow{x}\right)  \right]   &
=i\theta\varepsilon_{ab}\frac{\Omega}{\left(  2\pi\right)  ^{3}}%
\text{\ \ \ }a,b=1,2\label{eqn1ad}\\
\left[  \widehat{\varphi}_{a}\left(  t,\overrightarrow{x}\right)
,\widehat{\varphi}_{b}\left(  t,\overrightarrow{y}\right)  \right]   &
=0\text{ \ \ if }\overrightarrow{x}\neq\overrightarrow{y}\nonumber
\end{align}

In the limit $\Lambda\rightarrow\infty$, the noncommutative commutation
relations $\left(  \ref{eqn1ad}\right)  $ can be written as%

\begin{equation}
\left[  \widehat{\varphi}_{a}\left(  t,\overrightarrow{x}\right)
,\widehat{\varphi}_{b}\left(  t,\overrightarrow{y}\right)  \right]
=i\theta\varepsilon_{ab}\delta^{3}\left(  \overrightarrow{x}-\overrightarrow
{y}\right)
\end{equation}

The noncommutative canonical conjugates $\widehat{\pi}_{a}$ are defined by%

\begin{equation}
\widehat{\pi}_{a}\left(  x\right)  =\frac{\sqrt{\Omega}}{\left(  2\pi\right)
^{\frac{3}{2}}}\frac{\omega_{ab}^{-1}\left(  x\right)  }{\sqrt{\det\omega
^{-1}\left(  x\right)  }}\widehat{p}_{b,x}%
\end{equation}

Using the noncommutative relations $\left(  \ref{eqn1ac}\right)  ,$ one can
see that, in the limit $\Lambda\rightarrow\infty$, the noncommutative
canonical variables $\widehat{\varphi}_{a}$ and the noncommutative canonical
conjugates $\widehat{\pi}_{a}$ satisfy the noncommutative commutation relations%

\begin{align}
\left[  \widehat{\varphi}_{a}\left(  t,\overrightarrow{x}\right)
,\widehat{\pi}_{b}\left(  t,\overrightarrow{y}\right)  \right]   &
=i\delta^{3}\left(  \overrightarrow{x}-\overrightarrow{y}\right)  \delta
_{ab}\label{eqn6}\\
\left[  \widehat{\varphi}_{a}\left(  t,\overrightarrow{x}\right)
,\widehat{\varphi}_{b}\left(  t,\overrightarrow{y}\right)  \right]   &
=i\theta\varepsilon_{ab}\delta^{3}\left(  \overrightarrow{x}-\overrightarrow
{y}\right) \nonumber\\
\left[  \widehat{\pi}_{a}\left(  t,\overrightarrow{x}\right)  ,\widehat{\pi
}_{b}\left(  t,\overrightarrow{y}\right)  \right]   &  =0\nonumber
\end{align}

By generalizing the noncommutative harmonic oscillator construction\ an
extension of quantum field theory, based on the concept of noncommutative
fields satisfying the noncommutative commutation relations $\left(
\ref{eqn6}\right)  ,$ has been proposed in \cite{OSC13}-\cite{OSC15}, where
the properties and phenomenological implications of the noncommutative field
have been studied and applied to different problems including scalar, gauge
and fermionic fields \cite{OSC13}-\cite{OSC17}. Another approach, based on the
relations $\left(  \ref{eqn9}\right)  $ between the noncommutative variables
$\widehat{\varphi}_{a}$ and $\widehat{\pi}_{a}$ and the canonical variables
$\varphi_{a}$ and $\pi_{a}$, has been used and developed in the work of
Balachandran et al. \cite{bala 1}, where some periodic boundary conditions are
used to study a free massless scalar field in the noncommutative target space
$\mathbb{R}^{2}$, the theory has been quantized via the Hamiltonian formalism
and applied to the study of the deformed black radiation spectrum.

Our approach is different from that proposed in \cite{OSC13}-\cite{OSC15}, but
it bears some similarities with the approach used in the work of Balachandran
et al. \cite{bala 1}, both approaches are based on the relations $\left(
\ref{eqn9}\right)  $ between the noncommutative variables $\widehat{\varphi
}_{a}$ and $\widehat{\pi}_{a}$ and the canonical variables $\varphi_{a}$ and
$\pi_{a}$, but in our approach the equations of motions of the deformed theory
will be used to quantize the deformed theory via the Peierls bracket.

The noncommutative Hamiltonian density is assumed to have the form%

\begin{equation}
\widehat{\mathcal{H}}\mathcal{=}\frac{1}{2}\left(  \widehat{\pi}_{a}\right)
^{2}+\frac{1}{2}\left(  \overrightarrow{\nabla}\widehat{\varphi}_{a}\right)
^{2}+\frac{1}{2}\mu^{2}\left[  \widehat{\varphi}\right]  \left(
\widehat{\varphi}_{a}\right)  ^{2} \label{eqn8}%
\end{equation}

It is easy to see that the noncommutative commutation relations $\left(
\ref{eqn6}\right)  $ can be mapped to the canonical commutation relations
$\left(  \ref{eqn5}\right)  $ if the noncommutative variables $\widehat
{\varphi}_{a}$ and $\widehat{\pi}_{a}$ are related to the canonical variables
$\varphi_{a} $ and $\pi_{a}$ by the relations%

\begin{align}
\widehat{\varphi}_{a}  &  =\varphi_{a}-\frac{1}{2}\theta\varepsilon_{ab}%
\pi_{b}\label{eqn9}\\
\widehat{\pi}_{a}  &  =\pi_{a}\nonumber
\end{align}

Using these transformations, the noncommutative Hamiltonian density eq$\left(
\ref{eqn8}\right)  $ can be rewritten, up to a total derivative term and up to
second order in the parameter $\theta$, as%

\begin{multline}
\widehat{\mathcal{H}}\mathcal{=}\frac{1}{2}\pi^{\sim}\left(  \mathbb{I+}%
\frac{1}{4}\theta^{2}\left(  m^{2}-g\varepsilon\widehat{\sigma}\varepsilon
\right)  \right)  \pi-\frac{1}{8}\theta^{2}\pi^{\sim}\overrightarrow
{\mathbb{\nabla}}^{2}\pi+\frac{1}{2}\theta\pi^{\sim}\left(  m^{2}%
-\overrightarrow{\mathbb{\nabla}}^{2}+g\left(  \varphi_{a}\right)
^{2}\right)  \varepsilon\mathbb{\varphi}\label{eqn10}\\
\mathbb{+}\frac{1}{2}\varphi^{\sim}\left(  m^{2}-\overrightarrow
{\mathbb{\nabla}}^{2}+\frac{1}{2}g\left(  \varphi_{a}\right)  ^{2}\right)
\varphi+O\left(  \theta^{3}\right) \nonumber
\end{multline}

where%

\begin{equation}
\widehat{\sigma}_{ab}=\frac{\delta^{2}}{\delta\varphi_{a}\delta\varphi_{b}%
}\left[  \frac{1}{4}\left(  \varphi^{\sim}\varphi\right)  ^{2}\right]
=\varphi^{\sim}\varphi\delta_{ab}+2\varphi_{a}\varphi_{b}%
\end{equation}

with $\mathbb{I}$\ denotes the $2\times2$ unit matrix, and $\varphi^{\sim}$
denotes the transpose of $\mathbb{\varphi}.$

From now on we keep only the modifications due to the noncommutativity up to
second order in the parameter $\theta.$

The relation between $\pi_{a}$ and $\overset{\cdot}{\varphi}_{a}$ is given by%

\begin{equation}
\overset{\cdot}{\varphi}_{a}\left(  x\right)  =\frac{\delta\widehat{H}}%
{\delta\pi_{a}\left(  x\right)  } \label{eqn12}%
\end{equation}

where $\widehat{H}=\int d^{3}x\widehat{\mathcal{H}}.$ Using the expression of
$\widehat{\mathcal{H}}$ , one gets%

\begin{equation}
\overset{\cdot}{\varphi}_{a}\left(  x\right)  =\left(  \mathbb{I+}\frac{1}%
{4}\theta^{2}\left(  m^{2}-g\varepsilon\widehat{\sigma}\varepsilon\right)
\right)  _{ab}\pi_{b}\left(  x\right)  -\frac{1}{4}\theta^{2}\overrightarrow
{\mathbb{\nabla}}^{2}\pi_{a}\left(  x\right)  +\frac{1}{2}\theta\left(
m^{2}-\overrightarrow{\mathbb{\nabla}}^{2}+g\left(  \varphi_{a}\right)
^{2}\right)  \varepsilon_{ab}\mathbb{\varphi}_{b}\left(  x\right)
\label{eqn13}%
\end{equation}

From this relation we get, by iteration, the following expression of $\pi_{a}$%

\begin{equation}
\pi_{a}=\overset{\cdot}{\varphi}_{a}\mathbb{+}\frac{1}{4}\theta^{2}\left(
\overrightarrow{\mathbb{\nabla}}^{2}-\left(  m^{2}-g\varepsilon\widehat
{\sigma}\varepsilon\right)  \right)  _{ab}\overset{\cdot}{\varphi}_{b}%
-\frac{1}{2}\theta\left(  m^{2}-\overrightarrow{\mathbb{\nabla}}^{2}+g\left(
\varphi_{a}\right)  ^{2}\right)  \varepsilon_{ab}\mathbb{\varphi}_{b}
\label{eqn16}%
\end{equation}

We note that the noncommutative Hamiltonian density can be derived from the
following noncommutative Lagrangian density%

\begin{multline}
\widehat{\mathfrak{L}}=\frac{1}{2}\overset{\cdot}{\varphi^{\sim}}\left(
\mathbb{I+}\frac{1}{4}\theta^{2}\left(  \overrightarrow{\mathbb{\nabla}}%
^{2}-\left(  m^{2}-g\varepsilon\widehat{\sigma}\varepsilon\right)  \right)
\right)  \overset{\cdot}{\varphi}+\frac{1}{2}\theta\mathbb{\varphi}^{\sim
}\left(  m^{2}-\overrightarrow{\mathbb{\nabla}}^{2}+g\left(  \varphi
_{a}\right)  ^{2}\right)  \varepsilon\overset{\cdot}{\varphi}\label{eqn17}\\
-\frac{1}{2}\mathbb{\varphi}^{\sim}\left(  m^{2}-\overrightarrow
{\mathbb{\nabla}}^{2}+\frac{1}{2}g\left(  \varphi_{a}\right)  ^{2}%
\mathbb{-}\frac{1}{4}\theta^{2}\left(  m^{2}-\overrightarrow{\mathbb{\nabla}%
}^{2}+g\left(  \varphi_{a}\right)  ^{2}\right)  ^{2}\right)  \varphi
\end{multline}

via the usual Legendre transformation $\widehat{\mathfrak{L}}=\pi_{a}%
\overset{\cdot}{\varphi}_{a}-$ $\widehat{\mathcal{H}}$.

\section{Noncommutative field Equations}

Let us now consider the free theory, $g=0$, the noncommutative free
Hamiltonian density reads%

\begin{equation}
\widehat{\mathcal{H}}\mathcal{=}\frac{1}{2}\left(  1\mathbb{+}\frac{1}%
{4}\theta^{2}m^{2}\right)  \pi^{\sim}\pi-\frac{1}{8}\theta^{2}\pi^{\sim
}\overrightarrow{\mathbb{\nabla}}^{2}\pi+\frac{1}{2}\theta\pi^{\sim}\left(
m^{2}-\overrightarrow{\mathbb{\nabla}}^{2}\right)  \varepsilon\mathbb{\varphi
+}\frac{1}{2}\varphi^{\sim}\left(  m^{2}-\overrightarrow{\mathbb{\nabla}}%
^{2}\right)  \varphi\label{eqn18}%
\end{equation}

The noncommutative field equations are given by%

\begin{align}
\overset{\cdot}{\varphi}_{a}\left(  x\right)   &  =\frac{\delta\widehat{H}%
}{\delta\pi_{a}\left(  x\right)  }\label{eqn20}\\
\overset{\cdot}{\pi}_{a}\left(  x\right)   &  =-\frac{\delta\widehat{H}%
}{\delta\varphi_{a}\left(  x\right)  } \label{eqn20a}%
\end{align}

From the first equation we get%
\begin{equation}
\pi_{a}=\overset{\cdot}{\varphi}_{a}\mathbb{+}\frac{1}{4}\theta^{2}\left(
m^{2}-\overrightarrow{\mathbb{\nabla}}^{2}\right)  \overset{\cdot}{\varphi
}_{a}-\frac{1}{2}\theta\left(  m^{2}-\overrightarrow{\mathbb{\nabla}}%
^{2}\right)  \varepsilon_{ab}\mathbb{\varphi}_{b} \label{eqn21}%
\end{equation}

The second equation gives%
\begin{equation}
\overset{\cdot}{\pi}_{a}=-\left(  m^{2}-\overrightarrow{\mathbb{\nabla}}%
^{2}\right)  \varphi_{a}+\frac{1}{2}\theta\left(  m^{2}-\overrightarrow
{\mathbb{\nabla}}^{2}\right)  \varepsilon_{ab}\mathbb{\pi}_{b} \label{eqn22}%
\end{equation}

The noncommutative field equations eq$\left(  \ref{eqn21}\right)  $ and
eq$\left(  \ref{eqn22}\right)  $ may be written in the form%

\begin{equation}
\left[  -\mathcal{A}\frac{\partial^{2}}{\partial t^{2}}+\mathcal{B}%
\frac{\partial}{\partial t}-\mathcal{C}\right]  \varphi\left(  x\right)  =0
\label{eqn23a}%
\end{equation}

where%
\begin{align}
\mathcal{A}  &  \mathcal{=}\left[  1-\frac{1}{4}\theta^{2}\left(
m^{2}-\overrightarrow{\mathbb{\nabla}}^{2}\right)  \right]  \mathbb{I}%
=\mathcal{A}^{\sim}\label{eqn231c}\\
\mathcal{C}  &  \mathcal{=}\left[  1-\frac{1}{4}\theta^{2}\left(
m^{2}-\overrightarrow{\mathbb{\nabla}}^{2}\right)  \right]  \left(
m^{2}-\overrightarrow{\mathbb{\nabla}}^{2}\right)  \mathbb{I}=\mathcal{C}%
^{\sim}\nonumber\\
\mathcal{B}  &  \mathcal{=}\theta\left(  m^{2}-\overrightarrow{\mathbb{\nabla
}}^{2}\right)  \varepsilon=-\mathcal{B}^{\sim}\nonumber
\end{align}

and $\mathbb{A}^{\sim}$ denotes the transpose of the operator $\mathbb{A}.$

It is easy to see that the field equations eq$\left(  \ref{eqn23a}\right)  $
may be derived from the Lagrangian eq$\left(  \ref{eqn17}\right)  $%

\begin{equation}
\widehat{\mathbb{L}}=\int d^{3}\overrightarrow{x}\left[  \frac{1}{2}%
\overset{\cdot}{\varphi^{\sim}}\mathcal{A}\overset{\cdot}{\varphi}+\frac{1}%
{2}\mathbb{\varphi}^{\sim}\mathcal{B}\overset{\cdot}{\varphi}-\frac{1}%
{2}\mathbb{\varphi}^{\sim}\mathcal{C}\varphi\right]
\end{equation}

The general solution of eq$\left(  \ref{eqn23a}\right)  $ may be written as
(see Appendix A):%

\begin{equation}
\varphi\left(  x\right)  =\sum_{A}\left[  u_{A}^{\left(  +\right)  }\left(
x\right)  a_{A}+\overline{u}_{A}^{\left(  +\right)  }\left(  x\right)
\overline{a}_{A}\right]  +\sum_{A}\left[  u_{A}^{\left(  -\right)  }\left(
x\right)  b_{A}+\overline{u}_{A}^{\left(  -\right)  }\left(  x\right)
\overline{b}_{A}\right]  \label{eqn302}%
\end{equation}

for some time independent complex numbers $a_{A}$, $b_{A}$ and their complex
conjugates $\overline{a}_{A}$, $\overline{b}_{A},$ where $\overline{u}%
_{A}^{\left(  \pm\right)  }$ are the complex conjugates of the mode functions
$u_{A}^{\left(  \pm\right)  }$%

\begin{equation}
u_{A}^{\left(  \pm\right)  }\left(  t,\overrightarrow{x}\right)  =\chi
_{A}\left(  \overrightarrow{x}\right)  e^{-i\omega_{A}^{\left(  \pm\right)
}t}\zeta_{A}^{\left(  \pm\right)  }%
\end{equation}

The frequencies $\omega_{A}^{\left(  \pm\right)  }$ are given by eq$\left(
\ref{eqn30}\right)  $
\begin{equation}
\omega_{A}^{\left(  \pm\right)  }=\frac{1}{2}\left[  \mp\theta\overline
{\sigma}_{A}+\sqrt{4\overline{\sigma}_{A}+\theta^{2}\overline{\sigma}_{A}^{2}%
}\right]  \simeq\sqrt{\overline{\sigma}_{A}}\mp\frac{1}{2}\theta
\overline{\sigma}_{A}+\frac{1}{8}\theta^{2}\overline{\sigma}_{A}^{\frac{3}{2}}%
\end{equation}

where $\chi_{A}$ are the eigenvectors of the operator $-\overrightarrow
{\mathbb{\nabla}}^{2}$ with eigenvalues $\sigma_{A}$, and $\overline{\sigma
}_{A}=m^{2}+\sigma_{A}.$

Quantization of the noncommutative complex scalar field theory is
straightforward via the Peierls bracket (see \cite{Witt} for more details). In
the quantum theory, the field $\varphi$ becomes a Hermitian operator, and the
operator version of eq$\left(  \ref{eqn302}\right)  $%

\begin{equation}
\varphi\left(  x\right)  =\sum_{A}\left[  u_{A}^{\left(  +\right)  }\left(
x\right)  a_{A}+\overline{u}_{A}^{\left(  +\right)  }\left(  x\right)
a_{A}^{\ast}\right]  +\sum_{A}\left[  u_{A}^{\left(  -\right)  }\left(
x\right)  b_{A}+\overline{u}_{A}^{\left(  -\right)  }\left(  x\right)
b_{A}^{\ast}\right]  \label{eqn303}%
\end{equation}

holds for some constant operators $a_{A}$, $b_{A}$ and their Hermitian
conjugates $a_{A}^{\ast}$, $b_{A}^{\ast}$. By using the Wronskian relations
eq$\left(  \ref{eqn307}\right)  -$eq$\left(  \ref{eqn3010}\right)  $ we get%

\begin{align}
a_{A}  &  =-i\int d^{3}\overrightarrow{x}u_{A}^{\left(  +\right)  \ast}\left(
x\right)  \overleftrightarrow{\mathbb{W}}\varphi\left(  x\right)  \text{
\ \ \ }a_{A}^{\ast}=+\ i\int d^{3}\overrightarrow{x}u_{A}^{\left(  +\right)
\sim}\left(  x\right)  \overleftrightarrow{\mathbb{W}}\varphi\left(  x\right)
\label{eqn3011}\\
b_{A}  &  =-i\int d^{3}\overrightarrow{x}u_{A}^{\left(  -\right)  \ast}\left(
x\right)  \overleftrightarrow{\mathbb{W}}\varphi\left(  x\right)  \text{
\ \ \ }b_{A}^{\ast}=+\ i\int d^{3}\overrightarrow{x}u_{A}^{\left(  -\right)
\sim}\left(  x\right)  \overleftrightarrow{\mathbb{W}}\varphi\left(  x\right)
\nonumber
\end{align}

The quantum theory is obtained by setting
\begin{equation}
\left[  \varphi_{a}\left(  x\right)  ,\varphi_{b}\left(  y\right)  \right]
=i\widetilde{G}_{ab}\left(  x,y\right)  \label{eqn304}%
\end{equation}

where $\widetilde{G}$ is the commutator matrix%

\begin{multline}
\widetilde{G}\left(  x,y\right)  =-i\sum_{A}u_{A}^{\left(  +\right)  }\left(
x\right)  u_{A}^{\left(  +\right)  \ast}\left(  y\right)  +i\sum_{A}%
\overline{u}_{A}^{\left(  +\right)  }\left(  x\right)  u_{A}^{\left(
+\right)  \sim}\left(  y\right) \label{eqn305}\\
-i\sum_{A}u_{A}^{\left(  -\right)  }\left(  x\right)  u_{A}^{\left(  -\right)
\ast}\left(  y\right)  +i\sum_{A}\overline{u}_{A}^{\left(  -\right)  }\left(
x\right)  u_{A}^{\left(  -\right)  \sim}\left(  y\right) \nonumber
\end{multline}

Using the Wronskian relations eq$\left(  \ref{eqn307}\right)  -$eq$\left(
\ref{eqn3010}\right)  $ one can see that the commutator matrix $\widetilde{G}$
is the unique function that solves the Cauchy problem:
\begin{equation}
\varphi\left(  x\right)  =\int d^{3}\overrightarrow{y}\widetilde{G}\left(
x,y\right)  \overleftrightarrow{\mathbb{W}}\left(  y\right)  \varphi\left(
y\right)  \text{ \ \ at the same time }t=x^{0}=y^{0} \label{eqn3012}%
\end{equation}

Moreover the commutator matrix $\widetilde{G}$\ satisfies the equation%
\begin{equation}
\left[  -\mathcal{A}\frac{\partial^{2}}{\partial t^{2}}+\mathcal{B}%
\frac{\partial}{\partial t}-\mathcal{C}\right]  \widetilde{G}\left(
x,y\right)  =0 \label{eqn3013}%
\end{equation}

Using eq$\left(  \ref{eqn304}\right)  $ and the Wronskian relations eq$\left(
\ref{eqn307}\right)  -$eq$\left(  \ref{eqn3010}\right)  $ we get the
commutation relations%

\begin{gather}
\left[  a_{A},a_{B}^{\ast}\right]  =\delta_{AB}\ \ \ ,\ \ \ \left[
a_{A},a_{B}\right]  =\ \left[  a_{A}^{\ast},a_{B}^{\ast}\right]
=0\ \ \ \ \ \ \label{eqn3013a}\\
\left[  b_{A},b_{B}^{\ast}\right]  =\delta_{AB}\ \ \ ,\ \ \ \left[
b_{A},b_{B}\right]  =\ \left[  b_{A}^{\ast},b_{B}^{\ast}\right]
=0\ \ \ \nonumber\\
\left[  a_{A},b_{B}^{\ast}\right]  =\left[  a_{A},b_{B}\right]  =\ \left[
a_{A}^{\ast},b_{B}^{\ast}\right]  =\ \left[  a_{A}^{\ast},b_{B}\right]
=0\ \ \ \ \ \nonumber
\end{gather}

It is easy to show, by substituting the expression of $\pi$ eq$\left(
\ref{eqn21}\right)  $ into eq$\left(  \ref{eqn18}\right)  $, that the
noncommutative Hamiltonian operator can be written as%
\begin{equation}
\widehat{H}=\frac{1}{2}\int d^{3}\overrightarrow{x}\varphi^{\sim}\left(
x\right)  \overleftrightarrow{\mathbb{W}}\left(  x\right)  \overset{\cdot
}{\varphi}\left(  x\right)
\end{equation}

using the Wronskian relations eq$\left(  \ref{eqn307}\right)  -$eq$\left(
\ref{eqn3010}\right)  $, and the expression of $\mathbb{\varphi}$\ eq$\left(
\ref{eqn303}\right)  $, the noncommutative Hamiltonian operator $\widehat{H}$
of the system can be expressed as%
\begin{equation}
\widehat{H}=\sum_{A}\left(  \omega_{A}^{\left(  +\right)  }a_{A}^{\ast}%
a_{A}+\omega_{A}^{\left(  -\right)  }b_{A}^{\ast}b_{A}\right)  +\frac{1}%
{2}\sum_{A}\left(  \omega_{A}^{\left(  +\right)  }+\omega_{A}^{\left(
-\right)  }\right)  \label{eqn31}%
\end{equation}

where the commutation relations eq$\left(  \ref{eqn3013a}\right)  $ have been
used to get this form.

The noncommutative vacuum energy $E_{vac}$ reads%

\begin{equation}
E_{vac}=\left\langle vac\right\vert \widehat{H}\left\vert vac\right\rangle
=\frac{1}{2}\sum_{A}\left(  \omega_{A}^{\left(  +\right)  }+\omega
_{A}^{\left(  -\right)  }\right)  =\sum_{A}\left(  \sqrt{m^{2}+\sigma_{A}%
}+\frac{1}{8}\theta^{2}\left(  m^{2}+\sigma_{A}\right)  ^{\frac{3}{2}}\right)
\label{eqn32}%
\end{equation}

where the summation over $A$ is constrained by the condition eq$\left(
\ref{eqn3015}\right)  $. The noncommutative vacuum energy $E_{vac}$, in the
case were the free scalar field is confined in a D-dimensional rectangular box
of volume $V$ $=L^{D}$ with periodic boundary conditions on the walls of the
box, can be written as%
\begin{equation}
E_{vac}=\sum_{n_{1},n_{2},...,n_{D}}\left(  \left[  m^{2}+\sum_{k=1}%
^{D}\left(  \frac{2\pi n_{k}}{L}\right)  ^{2}\right]  ^{\frac{1}{2}}+\frac
{1}{8}\theta^{2}\left[  m^{2}+\sum_{k=1}^{D}\left(  \frac{2\pi n_{k}}%
{L}\right)  ^{2}\right]  ^{\frac{3}{2}}\right)  \label{eqn331}%
\end{equation}

where the summation over $n_{1},n_{2},...,n_{D}$ is constrained by the
condition eq$\left(  \ref{eqn3017}\right)  $.

In the limit $L\rightarrow\infty$ we can approximate the sums that occur in
eq$\left(  \ref{eqn331}\right)  $ with (divergent) integrals%
\begin{equation}
E_{vac}=V\int\frac{d^{D}\overrightarrow{p}}{\left(  2\pi\right)  ^{D}}\left(
\left[  \overrightarrow{p}^{2}+m^{2}\right]  ^{\frac{1}{2}}+\frac{1}{8}%
\theta^{2}\left[  \overrightarrow{p}^{2}+m^{2}\right]  ^{\frac{3}{2}}\right)
\end{equation}

Although these integrals are mathematically meaningless, one can use some sort
of regularization technique that makes the integrals finite. Using the
$\zeta-$function regularization (see the definitions and intermediate stages
of the calculation in Appendix B) \cite{TOMS}, we get the following expression
for the vacuum energy $E_{vac}$
\begin{equation}
E_{vac}=\left[  \frac{V\left[  m^{2}\right]  ^{\frac{D+1}{2}}}{\left(
4\pi\right)  ^{\frac{D}{2}}}\left[  l^{2}m^{2}\right]  ^{-\frac{s}{2}}%
\frac{\Gamma\left(  \frac{s-D-1}{2}\right)  }{\Gamma\left(  \frac{s-1}%
{2}\right)  }+\frac{1}{8}\theta^{2}\frac{V\left[  m^{2}\right]  ^{\frac
{D+3}{2}}}{\left(  4\pi\right)  ^{\frac{D}{2}}}\left[  l^{2}m^{2}\right]
^{-\frac{3s}{2}}\frac{\Gamma\left(  \frac{3s-D-3}{2}\right)  }{\Gamma\left(
\frac{3s-3}{2}\right)  }\right]  _{s\rightarrow0} \label{eqn332}%
\end{equation}

If $D$\ is even the right-hand side of eq$\left(  \ref{eqn332}\right)  $ is
analytic at $s=0$ with the result%

\begin{multline}
E_{vac}=\frac{V\left[  m^{2}\right]  ^{\frac{D+1}{2}}}{\left(  4\pi\right)
^{\frac{D}{2}}}\frac{\Gamma\left(  -\frac{D+1}{2}\right)  }{\Gamma\left(
-\frac{1}{2}\right)  }\left[  1+\frac{1}{8}\theta^{2}\frac{3m^{2}}{D+1}\right]
\\
=\frac{V\left[  m^{2}\right]  ^{\frac{D+1}{2}}}{\left(  4\pi\right)
^{\frac{D}{2}}}\frac{\left(  -2\right)  ^{\frac{D}{2}}}{1.3.5...\left(
D+1\right)  }\left[  1+\frac{1}{8}\theta^{2}\frac{3m^{2}}{D+1}\right]
\nonumber
\end{multline}

where we have used the following properties of the $\Gamma-$function
\cite{TOMS} \cite{Witt 2}%

\begin{align*}
z\Gamma\left(  z\right)   &  =\Gamma\left(  z+1\right) \\
\Gamma\left(  \frac{1}{2}-n\right)   &  =\frac{\left(  -2\right)  ^{n}%
\sqrt{\pi}}{1.3.5...\left(  2n-1\right)  }\text{ \ , \ }n=1,2,3,...
\end{align*}

When $D$ is odd the right-hand side of eq$\left(  \ref{eqn332}\right)  $ is
not analytic at $s=0,$ it has simple poles at $s=0,$ one simple pole from
$\Gamma\left(  \frac{s-D-1}{2}\right)  $ and another simple pole from
$\Gamma\left(  \frac{3s-D-3}{2}\right)  $. If we expand eq$\left(
\ref{eqn332}\right)  $ about the pole, in the case where $D$ $=3$, we find%
\begin{equation}
E_{vac}=-\frac{V}{2}\left(  \frac{m^{2}}{4\pi}\right)  ^{2}\left\{  \left[
1+\frac{m^{2}}{48}\theta^{2}\right]  \frac{2}{s}-\frac{1}{2}\left[
1+\frac{5m^{2}}{24}\theta^{2}\right]  -\left[  1+\frac{m^{2}}{16}\theta
^{2}\right]  \ln\frac{l^{2}m^{2}}{4}\right\}
\end{equation}

to get this expression the following formula has been used \cite{TOMS}
\cite{Witt 2}
\begin{equation}
\Gamma\left(  -n+\epsilon\right)  =\frac{\left(  -1\right)  ^{n}}{n!}\left(
\frac{1}{\epsilon}-\gamma+1+\frac{1}{2}+...+\frac{1}{n}\right)  +O\left(
\epsilon\right)
\end{equation}

where $n$ is a positive integer or zero, and $\gamma$ is the Euler constant.

The vacuum energy, when $D$ \ is odd, is divergent, this is just one example
of a variety of ultraviolet divergences that are encountered in quantum field
theory, they arise in a continuum theory due to the infinite number of degrees
of freedom that exist even in a finite volume, they can be reabsorbed into a
rescaling of the fields and into a rescaling of coupling constants. These
ultraviolet divergences can be eliminated by hand since only energy
differences can be observed, they are only important if we worry about
gravitational phenomena, since in general relativity any form of energy
contributes to the gravitational interaction \cite{WEINB} \cite{MAG}.

\section{Noncommutative Casimir Effect}

The Casimir effect is a non-classical electromagnetic, attractive force which
is concerned with vacuum fluctuations in the electromagnetic field between two
uncharged parallel conducting plates \cite{KAMilton}. The size of this force
was first predicted and calculated in 1948 by Casimir, who found that there is
an attractive force per unit area between two parallel, uncharged, perfectly
conducting plates separated by a distance $a$%
\[
F_{Casimir}=-\frac{\hbar c\pi^{2}}{240a^{4}}%
\]

This was first looked for by Sparnaay (1958), and recently has been confirmed
by Lamoreaux, Mohideen and Roy, and recently by Chan, Aksyuk, Kleiman, Bishop,
and Capasso \cite{KAMilton}\cite{casimir1}-\cite{casimir7}.

In this section we will consider the complex scalar field analogue of the
Casimir effect, for this we consider a massive complex scalar field in a
D-dimensional rectangular box, satisfying Dirichlet boundary conditions at
$x_{1}$ $=0$ and $x_{1}$ $=a$, but is unconfined in the remaining directions,
let $L_{1}=a$, $L_{2}=$ $L_{3}=...=L_{D}=$ $L$ be the sides of the box, and
$V=L_{1}L_{2}L_{3}$... $L_{D}$\ its volume, ultimately we will let $L$ becomes
infinitely large. The normalized eigenvectors $\chi_{A}$ and the eigenvalues
$\sigma_{A}$ of $-\overrightarrow{\mathbb{\nabla}}^{2}$ with Dirichlet
boundary conditions eq$\left(  \ref{eqn3017a}\right)  $ on the walls of the
box are given by eq$\left(  \ref{eqn3017b}\right)  $-eq$\left(  \ref{eqn3017c}%
\right)  $ \cite{TOMS}.

\bigskip The noncommutative vacuum energy $E_{vac}$ is given by eq$\left(
\ref{eqn32}\right)  $%

\begin{equation}
E_{vac}=E_{vac}^{\left(  C\right)  }+E_{vac}^{\left(  NC\right)  }=\sum
_{A}\left(  \sqrt{m^{2}+\sigma_{A}}+\frac{1}{8}\theta^{2}\left(  m^{2}%
+\sigma_{A}\right)  ^{\frac{3}{2}}\right)
\end{equation}

where $E_{vac}^{\left(  C\right)  }$ is the classical vacuum energy%

\begin{equation}
E_{vac}^{\left(  C\right)  }=\sum_{A}\sqrt{m^{2}+\sigma_{A}}=\sum_{n_{1}%
=1}^{\infty}\sum_{n_{2}=-\infty}^{\infty}...\sum_{n_{D}=-\infty}^{\infty}%
\sqrt{m^{2}+\left(  \frac{\pi n_{1}}{a}\right)  ^{2}+\sum_{k=2}^{D}\left(
\frac{2\pi n_{k}}{L}\right)  ^{2}} \label{eqn38}%
\end{equation}

and $E_{vac}^{\left(  NC\right)  }$ is the pure noncommutative vacuum energy%
\begin{align}
E_{vac}^{\left(  NC\right)  }  &  =\frac{1}{8}\theta^{2}\sum_{A}\left(
m^{2}+\sigma_{A}\right)  ^{\frac{3}{2}}\nonumber\\
&  =\frac{1}{8}\theta^{2}\sum_{n_{1}=1}^{\infty}\sum_{n_{2}=-\infty}^{\infty
}...\sum_{n_{D}=-\infty}^{\infty}\left[  m^{2}+\left(  \frac{\pi n_{1}}%
{a}\right)  ^{2}+\sum_{k=2}^{D}\left(  \frac{2\pi n_{k}}{L}\right)
^{2}\right]  ^{\frac{3}{2}} \label{eqn39}%
\end{align}

The classical vacuum energy can be written as%
\begin{equation}
E_{vac}^{\left(  C\right)  }=\underset{s\rightarrow0}{\lim}E\left(  s\right)
=E\left(  0\right)  \label{eqn41}%
\end{equation}

where the energy $\zeta-$function $E\left(  s\right)  $ is given by%

\begin{equation}
E\left(  s\right)  =l^{-s}\sum_{n_{1}=1}^{\infty}\sum_{n_{2}=-\infty}^{\infty
}...\sum_{n_{D}=-\infty}^{\infty}\left[  m^{2}+\left(  \frac{n_{1}\pi}%
{a}\right)  ^{2}+\sum_{k=2}^{D}\left(  \frac{2\pi n_{k}}{L}\right)
^{2}\right]  ^{\frac{1-s}{2}} \label{eqn42}%
\end{equation}

In the limit $L\rightarrow\infty$, we can replace the sums over $n_{2}%
,n_{3},...,n_{D}$ with integrals, so the energy $\zeta-$function becomes%

\begin{equation}
E\left(  s\right)  =l^{-s}\frac{V}{a}\sum_{n_{1}=1}^{\infty}\int\frac
{d^{D-1}\overrightarrow{p}}{\left(  2\pi\right)  ^{D-1}}\left[  \left(
\frac{n_{1}\pi}{a}\right)  ^{2}+\overrightarrow{p}^{2}+m^{2}\right]
^{\frac{1-s}{2}} \label{eqn43}%
\end{equation}

Using the relations eq$\left(  \ref{eqn46b}\right)  -$eq$\left(
\ref{eqn46a}\right)  $ in Appendix B, the energy $\zeta-$function, when
$m\rightarrow0$, can be written as%

\begin{equation}
E\left(  s\right)  =l^{-s}\frac{V}{\left(  4\pi\right)  ^{\frac{D-1}{2}}%
a}\left(  \frac{\pi}{a}\right)  ^{D-s}\frac{\Gamma\left(  \frac{s-D}%
{2}\right)  }{\Gamma\left(  \frac{s-1}{2}\right)  }\zeta\left(  s-D\right)
\label{eqn47a}%
\end{equation}

where $\zeta\left(  s\right)  =%
%TCIMACRO{\dsum \limits_{n=1}^{\infty}}%
%BeginExpansion
{\displaystyle\sum\limits_{n=1}^{\infty}}
%EndExpansion
n^{-s}$ is the Riemann $\zeta-$function.

Let us now consider the interesting case where $D=3,$ in this case the vacuum
energy takes the form%

\begin{equation}
E_{vac}^{\left(  C\right)  }=E\left(  0\right)  =-\frac{\pi^{2}V}{6a^{4}}%
\zeta\left(  -3\right)  =-\frac{\pi^{2}A}{720a^{3}}\label{eqn49}%
\end{equation}

where $A=L_{1}L_{2}=L^{2}$ is the area of the parallel $\left(
\text{uncharged conducting}\right)  $ plates.

The noncommutative vacuum energy $E_{vac}^{\left(  NC\right)  }$%

\begin{equation}
E_{vac}^{\left(  NC\right)  }=\left.  \frac{1}{8}\theta^{2}l^{-3s}\sum
_{n_{1}=1}^{\infty}\sum_{n_{2}=-\infty}^{\infty}...\sum_{n_{D}=-\infty
}^{\infty}\left[  m^{2}+\left(  \frac{n_{1}\pi}{a}\right)  ^{2}+\sum_{k=2}%
^{D}\left(  \frac{2\pi n_{k}}{L}\right)  ^{2}\right]  ^{\frac{3\left(
1-s\right)  }{2}}\right\vert _{s\rightarrow0} \label{eqn51}%
\end{equation}

\bigskip can be written as (see eq$\left(  \ref{eqn52a}\right)  -$eq$\left(
\ref{eqn52b}\right)  $ in Appendix B)%

\begin{equation}
E_{vac}^{\left(  NC\right)  }=\frac{1}{8}\theta^{2}l^{-3s}\frac{V}{a\left(
4\pi\right)  ^{\frac{D-1}{2}}}\frac{\Gamma\left(  \frac{\left(  3s-D-2\right)
}{2}\right)  }{\Gamma\left(  \frac{3s-3}{2}\right)  }\sum_{n=1}^{\infty
}\left[  \left(  \frac{n\pi}{a}\right)  ^{2}+m^{2}\right]  ^{-\frac{\left(
3s-D-2\right)  }{2}} \label{eqn55b}%
\end{equation}

When $m\rightarrow0$, the noncommutative vacuum energy $E_{vac}^{\left(
NC\right)  }$ becomes%

\begin{equation}
E_{vac}^{\left(  NC\right)  }=\frac{1}{8}\theta^{2}\frac{V}{a\left(
4\pi\right)  ^{\frac{D-1}{2}}}\left(  \frac{\pi}{a}\right)  ^{D+2}\frac
{\Gamma\left(  \frac{\left(  3s-D-2\right)  }{2}\right)  }{\Gamma\left(
\frac{3s-3}{2}\right)  }\left(  \frac{l\pi}{a}\right)  ^{-3s}\zeta\left(
3s-D-2\right)  \label{eqn56}%
\end{equation}

In the case where $D=3,$ $E_{vac}^{\left(  NC\right)  }$ takes the form%

\begin{equation}
E_{vac}^{\left(  NC\right)  }=-\frac{1}{8}\theta^{2}\frac{\pi^{4}V}{4a^{6}%
}\frac{2}{5}\zeta\left(  -5\right)  =+\frac{1}{8}\theta^{2}\frac{\pi^{4}%
A}{2520a^{5}} \label{eqn58}%
\end{equation}

The total vacuum energy $E_{vac}=E_{vac}^{\left(  C\right)  }+E_{vac}^{\left(
NC\right)  }$, is given by%

\begin{equation}
E_{vac}=E_{vac}^{\left(  C\right)  }+E_{vac}^{\left(  NC\right)  }%
=-\frac{\hbar c\pi^{2}A}{720a^{3}}\left(  1-\frac{\pi^{2}\theta^{2}}{28a^{2}%
}\right)  \label{eqn60}%
\end{equation}

The Casimir force reads%

\begin{equation}
F_{Casimir}=-\frac{\partial E_{vac}}{\partial a}=-\frac{\hbar c\pi^{2}%
A}{240a^{6}}\left(  a^{2}-\frac{5}{84}\pi^{2}\theta^{2}\right)  \label{eqn61}%
\end{equation}

where the first term represents the classical attractive Casimir force, while
the second term represents the noncommutative Casimir force, which is
repulsive. From eq$\left(  \ref{eqn61}\right)  $ we see that the total vacuum
energy $E_{vac}$ has a minimum at%

\begin{equation}
a_{\min}=\sqrt{\frac{5}{84}}\pi\theta\text{ \ ,\ \ }\theta\neq0 \label{eqn62}%
\end{equation}

At the equilibrium point $a_{\min}$, the total vacuum energy $E_{vac}$ takes
the value%

\begin{equation}
E_{vac}^{\min}=E_{vac}\left(  a_{\min}\right)  =-\frac{\hbar c\pi^{2}%
A}{720a_{\min}^{3}}\left(  1-\frac{1}{28}\frac{\pi^{2}\theta^{2}}{a_{\min}%
^{2}}\right)  =-\left(  3.\,\allowbreak849\,7\times10^{-28}%
%TCIMACRO{\unit{J}}%
%BeginExpansion
\operatorname{J}%
%EndExpansion%
%TCIMACRO{\unit{m}}%
%BeginExpansion
\operatorname{m}%
%EndExpansion
\right)  \frac{A}{\theta^{3}} \label{eqn63}%
\end{equation}

It is well known that the motion near the equilibrium may be approximately
described as harmonic oscillations, indeed near the equilibrium we may write
$a=a_{\min}+\delta$, expanding the total vacuum energy $E_{vac}$ in a Taylor
series
\begin{equation}
E_{vac}\left(  a\right)  =E_{vac}\left(  a_{\min}\right)  +E_{vac}^{\prime
}\left(  a_{\min}\right)  \delta+\frac{1}{2}E_{vac}^{\prime}\left(  a_{\min
}\right)  \delta^{2}+... \label{eqn63a}%
\end{equation}

we get%

\begin{equation}
E_{vac}\left(  a\right)  =-\frac{\hbar c\pi^{2}A}{720a^{3}}\left(  1-\frac
{1}{28}\frac{\pi^{2}\theta^{2}}{a^{2}}\right)  \simeq-\frac{\hbar c\pi^{2}%
A}{1800a_{\min}^{3}}+\frac{1}{2}\left(  \frac{\hbar c\pi^{2}A}{120a_{\min}%
^{5}}\right)  \delta^{2} \label{eqn64}%
\end{equation}

Hence the equation of motion near the equilibrium may be derived from the
following $\left(  \text{ harmonic oscillator\ }\right)  $ Lagrangian%

\begin{equation}
L=\frac{1}{2}\rho A\overset{\cdot}{\delta}^{2}-\frac{1}{2}\rho A\omega
^{2}\delta^{2} \label{eqn65}%
\end{equation}

where $\rho$ is the density of the parallel plate, and $\omega$ is the angular
frequency of vibration%
\begin{equation}
\omega=\sqrt{\frac{\hbar c\pi^{2}}{120\rho a_{\min}^{5}}}=\frac{\allowbreak
3.\,\allowbreak949\,9\times10^{-13}}{\sqrt{\rho}}\frac{\allowbreak1}%
{\theta^{\frac{5}{2}}} \label{eqn66}%
\end{equation}

\section{Conclusion}

Thought this work we have considered a noncommutative complex scalar field
theory with self interaction, by imposing non commutativity to the canonical
commutation relations. The noncommutative field equations are derived and
solved, the vacuum energy is calculated to the second order in the parameter
of non commutativity. As an example, we have considered the Casimir effect,
due to the zero point fluctuations of the noncommutative complex scalar field.
It turns out that in spite of its smallness, the noncommutativity gives rise
to a repulsive force at the microscopic level, leading to an effective Casimr
potential with a minimum at the point $a_{\min}=\sqrt{\frac{5}{84}}\pi\theta$.

\appendix

\section{Mode functions and wronskian relations}

The noncommutative field equations eq$\left(  \ref{eqn21}\right)  $ and
eq$\left(  \ref{eqn22}\right)  $ may be written in the form%

\begin{equation}
\left[  -\mathcal{A}\frac{\partial^{2}}{\partial t^{2}}+\mathcal{B}%
\frac{\partial}{\partial t}-\mathcal{C}\right]  \varphi\left(  x\right)  =0
\label{eqn23}%
\end{equation}

where%
\begin{align}
\mathcal{A}  &  \mathcal{=}\left[  1-\frac{1}{4}\theta^{2}\left(
m^{2}-\overrightarrow{\mathbb{\nabla}}^{2}\right)  \right]  \mathbb{I}%
=\mathcal{A}^{\sim}\label{eqn231}\\
\mathcal{C}  &  \mathcal{=}\left[  1-\frac{1}{4}\theta^{2}\left(
m^{2}-\overrightarrow{\mathbb{\nabla}}^{2}\right)  \right]  \left(
m^{2}-\overrightarrow{\mathbb{\nabla}}^{2}\right)  \mathbb{I}=\mathcal{C}%
^{\sim}\nonumber\\
\mathcal{B}  &  \mathcal{=}\theta\left(  m^{2}-\overrightarrow{\mathbb{\nabla
}}^{2}\right)  \varepsilon=-\mathcal{B}^{\sim}\nonumber
\end{align}

and $\mathbb{A}^{\sim}$ denotes the transpose of the operator $\mathbb{A}.$

To get the general solution of eq$\left(  \ref{eqn23}\right)  $ one begins by
looking for solutions of the form \cite{Witt}%

\begin{equation}
u_{A}\left(  t,\overrightarrow{x}\right)  =\chi_{A}\left(  \overrightarrow
{x}\right)  e^{-i\omega_{A}t}\zeta_{A} \label{eqn24}%
\end{equation}

known as mode functions, where $\chi_{A}$ are the eigenvectors of the operator
$-\overrightarrow{\mathbb{\nabla}}^{2}$ with eigenvalues $\sigma_{A}$ \
\begin{equation}
-\overrightarrow{\mathbb{\nabla}}^{2}\chi_{A}\left(  \overrightarrow
{x}\right)  =\sigma_{A}\chi_{A}\left(  \overrightarrow{x}\right)
\label{eqn25}%
\end{equation}

and $\zeta_{A}$ are $2\times1$ constant columns.

Insertion of eq$\left(  \ref{eqn24}\right)  $ into eq$\left(  \ref{eqn23}%
\right)  $ leads to the eigenvector-eigenvalue problem%

\begin{equation}
\left[  \left(  1-\frac{1}{4}\theta^{2}\overline{\sigma}_{A}\right)  \left(
\overline{\sigma}_{A}-\omega_{A}^{2}\right)  +i\theta\overline{\sigma}%
_{A}\varepsilon\omega_{A}\right]  \zeta_{A}=0 \label{eqn26}%
\end{equation}

where we have used the abbreviation $\overline{\sigma}_{A}=m^{2}+\sigma_{A}.$

This eigenvector-eigenvalue problem has a non trivial solution if and only if
the frequencies $\omega_{A}$ are roots of the equation%

\begin{equation}
\det\left[  \left(  1-\frac{1}{4}\theta^{2}\overline{\sigma}_{A}\right)
\left(  \overline{\sigma}_{A}-\omega_{A}^{2}\right)  +i\theta\overline{\sigma
}_{A}\varepsilon\omega_{A}\right]  =0 \label{eqn27}%
\end{equation}

which can be written in the equivalent form%
\begin{equation}
\left(  1-\frac{1}{4}\theta^{2}\overline{\sigma}_{A}\right)  ^{2}\left(
\overline{\sigma}_{A}-\omega_{A}^{2}\right)  ^{2}-\theta^{2}\overline{\sigma
}_{A}^{2}\omega_{A}^{2}=0 \label{eqn28}%
\end{equation}

Hence, the frequencies $\omega_{A}$ are the positive roots of the equations%
\begin{equation}
\omega_{A}^{2}\pm\theta\overline{\sigma}_{A}\omega_{A}-\overline{\sigma}_{A}=0
\label{eqn29}%
\end{equation}

The solutions are given by%
\begin{align}
\omega_{A}^{\left(  +\right)  }  &  =\frac{1}{2}\left[  -\theta\overline
{\sigma}_{A}+\sqrt{4\overline{\sigma}_{A}+\theta^{2}\overline{\sigma}_{A}^{2}%
}\right]  \simeq\sqrt{\overline{\sigma}_{A}}-\frac{1}{2}\theta\overline
{\sigma}_{A}+\frac{1}{8}\theta^{2}\overline{\sigma}_{A}^{\frac{3}{2}%
}\label{eqn30}\\
\omega_{A}^{\left(  -\right)  }  &  =\frac{1}{2}\left[  +\theta\overline
{\sigma}_{A}+\sqrt{4\overline{\sigma}_{A}+\theta^{2}\overline{\sigma}_{A}^{2}%
}\right]  \simeq\sqrt{\overline{\sigma}_{A}}+\frac{1}{2}\theta\overline
{\sigma}_{A}+\frac{1}{8}\theta^{2}\overline{\sigma}_{A}^{\frac{3}{2}}\nonumber
\end{align}

Because the mode functions%
\begin{equation}
u_{A}^{\left(  \pm\right)  }\left(  t,\overrightarrow{x}\right)  =\chi
_{A}\left(  \overrightarrow{x}\right)  e^{-i\omega_{A}^{\left(  \pm\right)
}t}\zeta_{A}^{\left(  \pm\right)  } \label{eqn301}%
\end{equation}

form a complete set the general solution of eq$\left(  \ref{eqn23}\right)  $
may be expanded in terms of them:%

\begin{equation}
\varphi\left(  x\right)  =\sum_{A}\left[  u_{A}^{\left(  +\right)  }\left(
x\right)  a_{A}+\overline{u}_{A}^{\left(  +\right)  }\left(  x\right)
\overline{a}_{A}\right]  +\sum_{A}\left[  u_{A}^{\left(  -\right)  }\left(
x\right)  b_{A}+\overline{u}_{A}^{\left(  -\right)  }\left(  x\right)
\overline{b}_{A}\right]  \label{eqn302a}%
\end{equation}

for some time independent complex numbers $a_{A}$, $b_{A}$ and their complex
conjugates $\overline{a}_{A}$, $\overline{b}_{A},$ where $\overline{u}%
_{A}^{\left(  \pm\right)  }$ are the complex conjugates of the mode functions
$u_{A}^{\left(  \pm\right)  }$. Starting from the equations satisfied by the
mode functions $u_{A}^{\left(  \pm\right)  }$%

\begin{equation}
\left[  -\mathcal{A}\frac{\partial^{2}}{\partial t^{2}}+\mathcal{B}%
\frac{\partial}{\partial t}-\mathcal{C}\right]  u_{A}^{\left(  \pm\right)
}\left(  x\right)  =0 \label{eqn306}%
\end{equation}

one can see, after some algebraic operations \cite{Witt}, that these mode
functions satisfy the Wronskian relations
\begin{gather}
-i\int d^{3}\overrightarrow{x}u_{A}^{\left(  +\right)  \ast}%
\overleftrightarrow{\mathbb{W}}u_{B}^{\left(  +\right)  }=\delta_{AB}\text{
\ \ \ \ \ \ \ \ \ }+\ i\int d^{3}\overrightarrow{x}u_{A}^{\left(  +\right)
\sim}\overleftrightarrow{\mathbb{W}}\overline{u}_{B}^{\left(  +\right)
}=\delta_{AB}\label{eqn307}\\
\text{\ }-i\int d^{3}\overrightarrow{x}u_{A}^{\left(  +\right)  \sim
}\overleftrightarrow{\mathbb{W}}u_{B}^{\left(  +\right)  }=0\text{
\ \ \ \ \ \ \ \ \ }+i\int d^{3}\overrightarrow{x}u_{A}^{\left(  +\right)
\ast}\overleftrightarrow{\mathbb{W}}\overline{u}_{B}^{\left(  +\right)
}=0\nonumber
\end{gather}

\begin{gather}
-i\int d^{3}\overrightarrow{x}u_{A}^{\left(  -\right)  \ast}%
\overleftrightarrow{\mathbb{W}}u_{B}^{\left(  -\right)  }=\delta_{AB}\text{
\ \ \ \ \ \ \ \ \ }+\ i\int d^{3}\overrightarrow{x}u_{A}^{\left(  -\right)
\sim}\overleftrightarrow{\mathbb{W}}\overline{u}_{B}^{\left(  -\right)
}=\delta_{AB}\label{eqn308}\\
\text{\ }-i\int d^{3}\overrightarrow{x}u_{A}^{\left(  -\right)  \sim
}\overleftrightarrow{\mathbb{W}}u_{B}^{\left(  -\right)  }=0\text{
\ \ \ \ \ \ \ \ \ }+i\int d^{3}\overrightarrow{x}u_{A}^{\left(  -\right)
\ast}\overleftrightarrow{\mathbb{W}}\overline{u}_{B}^{\left(  -\right)
}=0\nonumber
\end{gather}

\begin{gather}
-i\int d^{3}\overrightarrow{x}u_{A}^{\left(  +\right)  \ast}%
\overleftrightarrow{\mathbb{W}}u_{B}^{\left(  -\right)  }=0\text{
\ \ \ \ \ \ \ \ \ }+\ i\int d^{3}\overrightarrow{x}u_{A}^{\left(  +\right)
\sim}\overleftrightarrow{\mathbb{W}}\overline{u}_{B}^{\left(  -\right)
}=0\label{eqn309}\\
\text{\ \ }-i\int d^{3}\overrightarrow{x}u_{A}^{\left(  -\right)  \ast
}\overleftrightarrow{\mathbb{W}}u_{B}^{\left(  +\right)  }=0\text{
\ \ \ \ \ \ \ \ \ }+i\int d^{3}\overrightarrow{x}u_{A}^{\left(  -\right)
\sim}\overleftrightarrow{\mathbb{W}}\overline{u}_{B}^{\left(  +\right)
}=0\nonumber
\end{gather}

\begin{gather}
-i\int d^{3}\overrightarrow{x}u_{A}^{\left(  +\right)  \sim}%
\overleftrightarrow{\mathbb{W}}u_{B}^{\left(  -\right)  }=0\text{
\ \ \ \ \ \ \ \ \ }+i\int d^{3}\overrightarrow{x}u_{A}^{\left(  +\right)
\ast}\overleftrightarrow{\mathbb{W}}\overline{u}_{B}^{\left(  -\right)
}=0\label{eqn3010}\\
-i\int d^{3}\overrightarrow{x}u_{A}^{\left(  -\right)  \sim}%
\overleftrightarrow{\mathbb{W}}u_{B}^{\left(  +\right)  }=0\text{
\ \ \ \ \ \ \ \ \ }+i\int d^{3}\overrightarrow{x}u_{A}^{\left(  -\right)
\ast}\overleftrightarrow{\mathbb{W}}\overline{u}_{B}^{\left(  +\right)
}=0\nonumber
\end{gather}

where
\begin{equation}
\overleftrightarrow{\mathbb{W}}\left(  x\right)  =-\mathcal{A}\left(
x\right)  \frac{\overrightarrow{\partial}}{\partial t}+\mathcal{A}\left(
x\right)  \frac{\overleftarrow{\partial}}{\partial t}+\mathcal{B}\left(
x\right)
\end{equation}

is the Wronskian operator corresponding to the differential operator
\cite{Witt}%

\begin{equation}
\mathbb{F}=-\mathcal{A}\frac{\partial^{2}}{\partial t^{2}}+\mathcal{B}%
\frac{\partial}{\partial t}-\mathcal{C}%
\end{equation}

The Wronskian operator $\overleftrightarrow{\mathbb{W}}$ has the following
symmetry and reality properties:
\begin{equation}
\overleftrightarrow{\mathbb{W}}^{\sim}=-\overleftrightarrow{\mathbb{W}}\text{
\ , \ }\overleftrightarrow{\mathbb{W}}^{\ast}=-\overleftrightarrow{\mathbb{W}}%
\end{equation}

Here $\overline{\mathcal{O}},\mathcal{O}^{\ast}$and $\mathcal{O}^{\sim}$denote
the complex conjugate, the Hermitian conjugate and the transpose of the matrix
(or the operator) $\mathcal{O}$, respectively.

In order that these Wronskian relations must hold, the operators $\mathcal{A}
$ and $\mathcal{C}$ must be positive definite operators, but the eigenvalues
of the operators $\mathcal{A}$ and $\mathcal{C}$ are given by%

\begin{align}
\mathcal{A}u_{A}^{\left(  \pm\right)  }\left(  x\right)   &  =\left[
1-\frac{1}{4}\theta^{2}\left(  m^{2}-\overrightarrow{\mathbb{\nabla}}%
^{2}\right)  \right]  u_{A}^{\left(  \pm\right)  }\left(  x\right)  =\left(
1-\frac{1}{4}\theta^{2}\overline{\sigma}_{A}\right)  u_{A}^{\left(
\pm\right)  }\left(  x\right) \\
\mathcal{C}u_{A}^{\left(  \pm\right)  }\left(  x\right)   &  =\left[
1-\frac{1}{4}\theta^{2}\left(  m^{2}-\overrightarrow{\mathbb{\nabla}}%
^{2}\right)  \right]  \left(  m^{2}-\overrightarrow{\mathbb{\nabla}}%
^{2}\right)  u_{A}^{\left(  \pm\right)  }\left(  x\right)  =\left(  1-\frac
{1}{4}\theta^{2}\overline{\sigma}_{A}\right)  \overline{\sigma}_{A}%
u_{A}^{\left(  \pm\right)  }\left(  x\right) \nonumber
\end{align}

so these eigenvalues are not positive for all indices $A$, to solve this
problem we use the fact that $\theta\sim10^{-13}m$ \cite{Saha1}-\cite{Saha4},
so $\left(  1-\frac{1}{4}\theta^{2}\overline{\sigma}_{A}\right)  >0$ for all
indices $A$ such that $\overline{\sigma}_{A}<\frac{4}{\theta^{2}}\sim10^{26}$,
to make the spectrum of the operators $\mathcal{A}$ and $\mathcal{C}$
\ bounded we impose the following boundary conditions on the eigenfunctions
$\chi_{A}\left(  \overrightarrow{x}\right)  $ of the operator
$-\overrightarrow{\mathbb{\nabla}}^{2}$%
\begin{equation}
\left\vert \frac{\partial}{\partial x_{j}}\chi_{A}\left(  x_{1},...,x_{j}%
...,x_{D}\right)  \right\vert _{\overrightarrow{x}=\overrightarrow{a}}%
\leq\frac{\alpha}{\theta}\text{ \ \ \ \ }\ j=1,2,...,D \label{eqn3015}%
\end{equation}

at some arbitrary point $\overrightarrow{x}=\overrightarrow{a}$, and $\alpha$
is some constant with dimension $\left(  \text{length}\right)  ^{-\frac{3}{2}%
}.$ Note that in the classical limit where $\theta\rightarrow0$ this condition
is trivially satisfied.

As an example we consider the free scalar field confined in a D-dimensional
rectangular box of volume $V$ $=L^{D}$ and impose periodic boundary conditions
on the walls of the box, the normalized eigenfunctions $\chi_{A}\left(
\overrightarrow{x}\right)  $ of the operator $-\overrightarrow{\mathbb{\nabla
}}^{2}$, are \cite{TOMS}%

\begin{equation}
\sqrt{\frac{1}{V}}\exp\left[  \sum_{k=1}^{D}\frac{2\pi in_{k}}{L}x_{k}\right]
\ \ \ \ \text{with\ }\ \ n_{k}=0,\pm1,\pm2,...,\text{ for each }k=1,2,...,D
\label{eqn3016}%
\end{equation}

in this case the boundary conditions eq$\left(  \ref{eqn3015}\right)  $ read%

\begin{equation}
\left\vert \frac{2\pi n_{j}}{L}\right\vert \leq\frac{\alpha\sqrt{V}}{\theta
}\text{ \ \ \ \ \ \ \ \ \ \ \ \ \ \ \ \ \ \ \ \ \ \ \ \ \ \ \ \ \ }%
\ j=1,2,...,D \label{eqn3017}%
\end{equation}

if we choose $\alpha=\frac{1}{\sqrt{DV}}$we get%

\begin{equation}
\frac{1}{4}\theta^{2}\overline{\sigma}_{A}=\frac{1}{4}\theta^{2}\left[
m^{2}+\sum_{k=1}^{D}\left(  \frac{2\pi n_{k}}{L}\right)  ^{2}\right]
\leq\frac{1}{4}\theta^{2}m^{2}+\frac{\alpha^{2}DV}{4}<1
\end{equation}

where we have used the fact that $\theta$ is an infinitesimal parameter such
that $\theta^{2}m^{2}<1.$ Hence $\mathcal{A}$ and $\mathcal{C}$ are positive
definite operators.

As a second example, we consider the free scalar field confined in a
D-dimensional rectangular box of volume $V$ $=L^{D}$ and impose Dirichlet
boundary conditions on the walls of the box, the normalized eigenvectors
$\chi_{A}$ of $-\overrightarrow{\mathbb{\nabla}}^{2}$ with Dirichlet boundary
conditions on the walls of the box%

\begin{align}
\chi_{A}\left(  0,x_{2},x_{3},...,x_{D}\right)   &  =\chi_{A}\left(
L,x_{2},x_{3},...,x_{D}\right)  =0\label{eqn3017a}\\
\chi_{A}\left(  x_{1},...,x_{k-1},0,x_{k+1},...,x_{D}\right)   &  =\chi
_{A}\left(  x_{1},...,x_{k-1},L,x_{k+1},...,x_{D}\right)  \text{ \ ,
}k=2,...,D\nonumber
\end{align}

are given by \cite{TOMS}%

\begin{gather}
-\overrightarrow{\mathbb{\nabla}}^{2}\chi_{A}\left(  \overrightarrow
{x}\right)  =\sigma_{A}\chi_{A}\left(  \overrightarrow{x}\right)
\label{eqn3017b}\\
\chi_{A}\left(  \overrightarrow{x}\right)  =\sqrt{\frac{2}{V}}\sin\left(
\frac{\pi n_{1}}{L}x_{1}\right)  \exp\left[  \sum_{k=2}^{D}\frac{2\pi in_{k}%
}{L}x_{k}\right] \nonumber
\end{gather}

with $n_{1}=1,2,...$ and $n_{k}=0,\pm1,\pm2,...$ for $k=2,3,...,D$.

The eigenvalues are given by%

\begin{equation}
\sigma_{A}\equiv\sigma_{n_{1}n_{2}...n_{D}}=\left(  \frac{\pi n_{1}}%
{L}\right)  ^{2}+\sum_{k=2}^{D}\left(  \frac{2\pi n_{k}}{L}\right)  ^{2}
\label{eqn3017c}%
\end{equation}

in this case the boundary conditions eq$\left(  \ref{eqn3015}\right)  $ read%

\begin{gather}
\left\vert \frac{\partial}{\partial x_{j}}\chi_{A}\left(  x_{1},...,x_{j}%
...,x_{D}\right)  \right\vert _{\overrightarrow{x}=\overrightarrow{a}}%
=\sqrt{\frac{2}{V}}\left\vert \sin\left(  \frac{\pi n_{1}}{L}a_{1}\right)
\right\vert \left\vert \frac{2\pi n_{j}}{L}\right\vert \leq\frac{\alpha
}{\theta}\text{ \ \ ,\ \ \ \ \ \ \ }\ j=2,...,D\\
\left\vert \frac{\partial}{\partial x_{1}}\chi_{A}\left(  x_{1},...,x_{j}%
...,x_{D}\right)  \right\vert _{\overrightarrow{x}=\overrightarrow{a}}%
=\sqrt{\frac{2}{V}}\left\vert \frac{n_{1}\pi}{L}\cos\left(  \frac{\pi n_{1}%
}{L}a_{1}\right)  \right\vert \leq\frac{\alpha}{\theta}%
\end{gather}

leading to the constraints%

\begin{align}
\left\vert \frac{2\pi n_{j}}{L}\right\vert  &  \leq\frac{1}{\left\vert
\sin\left(  \frac{\pi n_{1}}{L}a_{1}\right)  \right\vert }\frac{\alpha\sqrt
{V}}{\sqrt{2}\theta}\text{ \ \ \ \ ,\ \ }\frac{L}{a_{1}}\notin\mathbb{N}%
\text{\ \ ,\ \ \ }j=1,2,...,D\\
\left\vert \frac{n_{1}\pi}{L}\right\vert  &  \leq\frac{1}{\left\vert
\cos\left(  \frac{\pi n_{1}}{L}a_{1}\right)  \right\vert }\frac{\alpha\sqrt
{V}}{\sqrt{2}\theta}\text{ \ \ \ \ ,\ \ }\frac{L}{a_{1}}\notin\mathbb{N}%
\end{align}

if we choose $\alpha=\sqrt{\frac{a_{1}}{L\left(  D-1\right)  V}}$ and
$a_{1}\approx0,$\ we get%

\begin{equation}
\frac{1}{4}\theta^{2}\overline{\sigma}_{A}=\frac{1}{4}\theta^{2}\left[
m^{2}+\left(  \frac{\pi n_{1}}{L}\right)  ^{2}+\sum_{k=2}^{D}\left(
\frac{2\pi n_{k}}{L}\right)  ^{2}\right]  \leq\frac{1}{4}\theta^{2}m^{2}%
+\frac{\left(  D-1\right)  L}{a_{1}}\frac{\alpha^{2}V}{8\pi}<1
\end{equation}

where we have used the fact that $\theta$ is an infinitesimal parameter such
that $\theta^{2}m^{2}<1.$ Hence $\mathcal{A}$ and $\mathcal{C}$ are positive
definite operators.

\section{Zeta function regularization}

The noncommutative vacuum energy $E_{vac}$ is given by%
\begin{equation}
E_{vac}=E_{vac}^{\left(  C\right)  }+E_{vac}^{\left(  NC\right)  }=\sum
_{A}\left(  \sqrt{m^{2}+\sigma_{A}}+\frac{1}{8}\theta^{2}\left(  m^{2}%
+\sigma_{A}\right)  ^{\frac{3}{2}}\right)
\end{equation}

where $E_{vac}^{\left(  C\right)  }$ is the classical vacuum energy%

\begin{equation}
E_{vac}^{\left(  C\right)  }=\sum_{A}\sqrt{m^{2}+\sigma_{A}}=\sum_{n_{1}%
=1}^{\infty}\sum_{n_{2}=-\infty}^{\infty}...\sum_{n_{D}=-\infty}^{\infty}%
\sqrt{m^{2}+\left(  \frac{\pi n_{1}}{a}\right)  ^{2}+\sum_{k=2}^{D}\left(
\frac{2\pi n_{k}}{L}\right)  ^{2}} \label{eqn38a}%
\end{equation}

and $E_{vac}^{\left(  NC\right)  }$ is the pure noncommutative vacuum energy%
\begin{equation}
E_{vac}^{\left(  NC\right)  }=\frac{1}{8}\theta^{2}\sum_{n_{1}=1}^{\infty}%
\sum_{n_{2}=-\infty}^{\infty}...\sum_{n_{D}=-\infty}^{\infty}\left[
m^{2}+\left(  \frac{\pi n_{1}}{a}\right)  ^{2}+\sum_{k=2}^{D}\left(
\frac{2\pi n_{k}}{L}\right)  ^{2}\right]  ^{\frac{3}{2}} \label{eqn39a}%
\end{equation}

To deal with the infinite sum of zero point energies in eq$\left(
\ref{eqn38a}\right)  $ and eq$\left(  \ref{eqn39a}\right)  $, we must
introduce a regularization method to extract finite expression \cite{TOMS}%
\cite{Dittrich}\cite{ZETA}\cite{Witt 2}. One elegant way for doing this is to
use $\zeta-$function regularization \cite{TOMS}, the idea of the method is to
define the divergent sum $\sum_{A}E_{A}$ over zero-point energies in
eq$\left(  \ref{eqn38a}\right)  $ and eq$\left(  \ref{eqn39a}\right)  $ by the
analytic continuation of a convergent sum. First, we consider the infinite sum
in eq$\left(  \ref{eqn38a}\right)  $, we define the energy $\zeta-$function by
\cite{TOMS}%

\begin{equation}
E\left(  s\right)  =\sum_{A}E_{A}\left(  lE_{A}\right)  ^{-s}%
\end{equation}

where $E_{A}=$ $\sqrt{m^{2}+\sigma_{A}},$ $s$ is a complex variable and $l$ is
a constant with units of length, introduced to keep$\left(  lE_{A}\right)  $
dimensionless. This ensures that $E\left(  s\right)  $ has dimensions of
energy for all values of $s$.

The classical vacuum energy can be written as%
\begin{equation}
E_{vac}^{\left(  C\right)  }=\underset{s\rightarrow0}{\lim}E\left(  s\right)
=E\left(  0\right)
\end{equation}

where the energy $\zeta-$function $E\left(  s\right)  $ is given by%

\begin{equation}
E\left(  s\right)  =l^{-s}\sum_{n_{1}=1}^{\infty}\sum_{n_{2}=-\infty}^{\infty
}...\sum_{n_{D}=-\infty}^{\infty}\left[  m^{2}+\left(  \frac{n_{1}\pi}%
{a}\right)  ^{2}+\sum_{k=2}^{D}\left(  \frac{2\pi n_{k}}{L}\right)
^{2}\right]  ^{\frac{1-s}{2}}%
\end{equation}

In the limit $L\rightarrow\infty$, we can replace the sums over $n_{2}%
,n_{3},...,n_{D}$ with integrals, so the energy $\zeta-$function becomes%

\begin{equation}
E\left(  s\right)  =l^{-s}\frac{V}{a}\sum_{n_{1}=1}^{\infty}\int\frac
{d^{D-1}\overrightarrow{p}}{\left(  2\pi\right)  ^{D-1}}\left[  \left(
\frac{n_{1}\pi}{a}\right)  ^{2}+\overrightarrow{p}^{2}+m^{2}\right]
^{\frac{1-s}{2}} \label{eqn46b}%
\end{equation}

Using the identity%

\begin{equation}
a^{-z}=\frac{1}{\Gamma\left(  z\right)  }\int_{0}^{\infty}dtt^{z-1}e^{-at}
\label{eqn1A}%
\end{equation}

which holds for $\operatorname{Re}(z)>0$ and $\operatorname{Re}(a)>0$, where
$\Gamma\left(  z\right)  $ is Gamma function%
\begin{equation}
\Gamma\left(  z\right)  =\int_{0}^{\infty}dtt^{z-1}e^{-t} \label{eqn44}%
\end{equation}

defined for $\operatorname{Re}(z)>0$, we obtain the following expression for
the energy $\zeta-$function%

\begin{equation}
E\left(  s\right)  =l^{-s}\frac{V}{a}\sum_{n_{1}=1}^{\infty}\frac{1}%
{\Gamma\left(  \frac{s-1}{2}\right)  }\int_{0}^{\infty}dtt^{\frac{s-3}{2}}%
\exp\left(  -\left[  \left(  \frac{n_{1}\pi}{a}\right)  ^{2}+m^{2}\right]
t\right)  \int\frac{d^{D-1}\overrightarrow{p}}{\left(  2\pi\right)  ^{D-1}%
}\exp\left(  -\overrightarrow{p}^{2}t\right) \nonumber
\end{equation}

the integration over the $\left(  D-1\right)  -$dimensional momentum integral
on the right-hand side can be performed with the help of the relations
\cite{QFT4} \cite{TOMS} \cite{Witt 2} \cite{QFT2}
\begin{equation}
\int d^{n}qf\left(  q^{2}\right)  =\frac{2\pi^{\frac{n}{2}}}{\Gamma\left(
\frac{n}{2}\right)  }\int_{0}^{+\infty}dkk^{n-1}f\left(  q^{2}\right)
\label{eqn45a}%
\end{equation}

and%

\begin{equation}
\int_{0}^{+\infty}dtt^{2s-1}e^{-\alpha t^{2}}=\frac{\alpha^{-s}}{2}%
\Gamma\left(  s\right)  \label{eqn45b}%
\end{equation}
with the results%

\begin{equation}
E\left(  s\right)  =l^{-s}\frac{V}{\left(  4\pi\right)  ^{\frac{D-1}{2}}%
a}\frac{\Gamma\left(  \frac{s-D}{2}\right)  }{\Gamma\left(  \frac{s-1}%
{2}\right)  }\sum_{n_{1}=1}^{\infty}\left[  \left(  \frac{n_{1}\pi}{a}\right)
^{2}+m^{2}\right]  ^{\frac{D-s}{2}} \label{eqn46}%
\end{equation}

When $m\rightarrow0$, the energy $\zeta-$function becomes%

\begin{equation}
E\left(  s\right)  =l^{-s}\frac{V}{\left(  4\pi\right)  ^{\frac{D-1}{2}}%
a}\left(  \frac{\pi}{a}\right)  ^{D-s}\frac{\Gamma\left(  \frac{s-D}%
{2}\right)  }{\Gamma\left(  \frac{s-1}{2}\right)  }\zeta\left(  s-D\right)
\label{eqn46a}%
\end{equation}

where $\zeta\left(  s\right)  =%
%TCIMACRO{\dsum \limits_{n=1}^{\infty}}%
%BeginExpansion
{\displaystyle\sum\limits_{n=1}^{\infty}}
%EndExpansion
n^{-s}$ is the Riemann $\zeta-$function.

By the same steps we will now calculate the noncommutative vacuum energy
$E_{vac}^{\left(  NC\right)  }$, let $\mathcal{E}\left(  s\right)  $ be the
energy $\zeta-$function%

\begin{equation}
\mathcal{E}\left(  s\right)  =l^{-3s}\sum_{n_{1}=1}^{\infty}\sum
_{n_{2}=-\infty}^{\infty}...\sum_{n_{D}=-\infty}^{\infty}\left[  m^{2}+\left(
\frac{n_{1}\pi}{a}\right)  ^{2}+\sum_{k=2}^{D}\left(  \frac{2\pi n_{k}}%
{L}\right)  ^{2}\right]  ^{\frac{3\left(  1-s\right)  }{2}} \label{eqn52a}%
\end{equation}

\bigskip then%
\begin{equation}
E_{vac}^{\left(  NC\right)  }=\underset{s\rightarrow0}{\lim\mathcal{E}\left(
s\right)  }=\mathcal{E}\left(  0\right)  \label{eqn52}%
\end{equation}

In the limit $L\rightarrow\infty$, we can replace the sums over $n_{2}%
,n_{3},...,n_{D}$ with integrals, so the energy $\zeta-$function becomes%

\begin{equation}
\mathcal{E}\left(  s\right)  =l^{-3s}\frac{V}{a}\sum_{n=1}^{\infty}\int
\frac{d^{D-1}\overrightarrow{p}}{\left(  2\pi\right)  ^{D-1}}\left[  \left(
\frac{n\pi}{a}\right)  ^{2}+\overrightarrow{p}^{2}+m^{2}\right]
^{\frac{3\left(  1-s\right)  }{2}} \label{eqn53}%
\end{equation}

using the relation$\left(  \ref{eqn1A}\right)  $, one gets%

\begin{equation}
\mathcal{E}\left(  s\right)  =l^{-3s}\frac{V}{a}\sum_{n=1}^{\infty}\frac
{1}{\Gamma\left(  \frac{3}{2}\left(  s-1\right)  \right)  }\int_{0}^{\infty
}dtt^{\frac{\left(  3s-5\right)  }{2}}\int\frac{d^{D-1}\overrightarrow{p}%
}{\left(  2\pi\right)  ^{D-1}}e^{-\left[  \left(  \frac{n\pi}{a}\right)
^{2}+\overrightarrow{p}^{2}+m^{2}\right]  t} \label{eqn54}%
\end{equation}

the integration over the $\left(  D-1\right)  -$dimensional momentum integral
on the right-hand side can be performed with the help of the relations
$\left(  \ref{eqn45a}\right)  $ and $\left(  \ref{eqn45b}\right)  $, one finds%

\begin{equation}
\mathcal{E}\left(  s\right)  =l^{-3s}\frac{V}{a\left(  4\pi\right)
^{\frac{D-1}{2}}}\frac{1}{\Gamma\left(  \frac{3}{2}\left(  s-1\right)
\right)  }\sum_{n=1}^{\infty}\left[  \left(  \frac{n\pi}{a}\right)  ^{2}%
+m^{2}\right]  ^{-\frac{\left(  3s-D-2\right)  }{2}}\int_{0}^{\infty
}dtt^{\frac{\left(  3s-D-2\right)  }{2}-1}e^{-t}\nonumber
\end{equation}

using eq$\left(  \ref{eqn44}\right)  $ to perform the integration over $t$, we get%

\begin{equation}
\mathcal{E}\left(  s\right)  =l^{-3s}\frac{V}{a\left(  4\pi\right)
^{\frac{D-1}{2}}}\frac{\Gamma\left(  \frac{\left(  3s-D-2\right)  }{2}\right)
}{\Gamma\left(  \frac{3s-3}{2}\right)  }\sum_{n=1}^{\infty}\left[  \left(
\frac{n\pi}{a}\right)  ^{2}+m^{2}\right]  ^{-\frac{\left(  3s-D-2\right)  }%
{2}} \label{eqn52b}%
\end{equation}

When $m\rightarrow0$, the energy $\zeta-$function becomes%

\begin{equation}
\mathcal{E}\left(  s\right)  =\frac{V}{a\left(  4\pi\right)  ^{\frac{D-1}{2}}%
}\left(  \frac{\pi}{a}\right)  ^{D+2}\frac{\Gamma\left(  \frac{\left(
3s-D-2\right)  }{2}\right)  }{\Gamma\left(  \frac{3s-3}{2}\right)  }\left(
\frac{l\pi}{a}\right)  ^{-3s}\zeta\left(  3s-D-2\right)
\end{equation}

\end{document}